\documentclass[a4paper]{article}

\usepackage{INTERSPEECH2022}

\usepackage{booktabs}
\usepackage{multirow}
\usepackage{tabularx}
\usepackage{enumerate}
\usepackage{amssymb}
\usepackage{color}
\usepackage{url}

\usepackage{enumitem}

\title{Leveraging Symmetrical  Convolutional Transformer  Networks for \\ Speech to Singing Voice Style Transfer} 
\name{ Shrutina Agarwal$^\$$, Sriram Ganapathy$^\$$, Naoya Takahashi$^*$}
\address{$^\$$LEAP lab, Indian Institute of Science, Bangalore, India.\\
$^*$Sony Group Corporation, Tokyo, Japan.
\email{sriramg@iisc.ac.in}}
\graphicspath{{images/}}
\begin{document}
\ninept
\fontsize{9.0pt}{11.2pt}\selectfont

\maketitle
\begin{abstract}
In this paper, we propose a model to perform style transfer of speech to singing voice. Contrary to the previous signal processing-based methods, which require high-quality singing templates or phoneme  synchronization, we explore a data-driven approach for  the problem of converting natural speech to singing voice. 
We develop a novel neural network architecture, called SymNet, which  models the  alignment of the input speech  with the target melody while preserving the speaker identity and naturalness. The proposed SymNet model is comprised of symmetrical stack of three types of layers - convolutional, transformer, and self-attention layers. 
The paper also explores novel data augmentation and generative loss annealing methods to facilitate the model training. Experiments are performed on the 
 NUS and NHSS datasets which consist of parallel data of speech and singing voice. In these experiments, we show that the proposed SymNet model improves the objective reconstruction quality significantly over the previously  published methods and  baseline architectures. Further, a subjective listening test  confirms the improved quality of the audio obtained  using the proposed approach (absolute improvement of $0.37$ in mean opinion score measure over the baseline system). 
\end{abstract}

\section{Introduction}
The task of speech to singing (STS) voice style transfer is the problem of converting the speech signal to a natural sounding singing voice. The inputs to the STS tasks are the  speech waveform and a target melody contour.
The key challenges involved in this problem relate to the ability of the STS system to generate natural voice which preserves the speaker and phonetic content while incorporating the melody.


The style transfer on other domains like images~\cite{gatys2016image} has been more successful than in audio and speech domain. 
In the recent past, several works attempt the goal of transforming   specific properties of audio like the speaker characteristics (referred to as “style”) while keeping linguistic characteristics intact (referred to as “content”). Haque et al.~\cite{relate1} proposed  a system to transform the timbre of the speech. Similarly, Mor et al.~\cite{relate2} proposed a translation framework across various musical instruments, styles and genres. 
Wu et al.~\cite{relate3} explored a cycle-generative adversarial network (cycle-GAN) based framework to perform singing voice conversion.  
In this paper, we propose a novel neural architecture for the task fo speech to singing voice conversion. 

\begin{figure*}[t!]
\begin{center}
\centerline{\includegraphics[width=2.05\columnwidth]{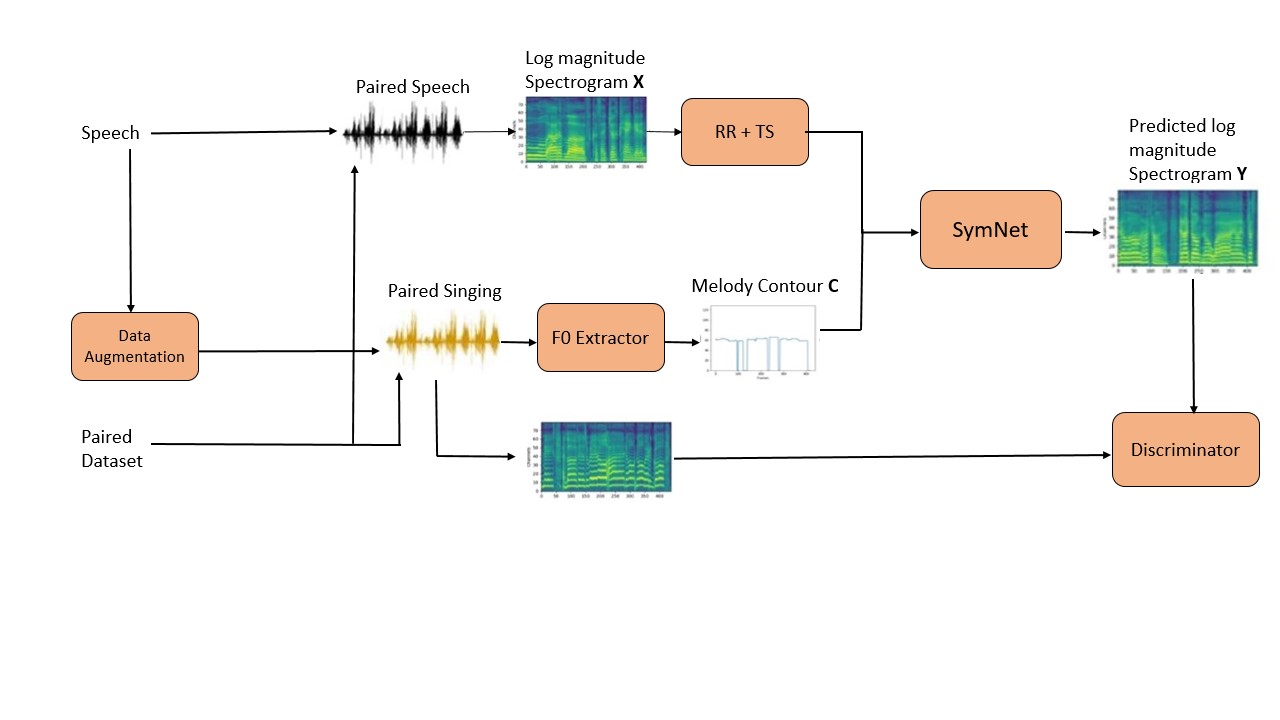}}
\vspace{-1.1in}
\caption{Block schematic of the proposed approach to STS. Here, RR is random sampling and TS is time-stretching.}
\vspace{-0.15in}
\label{fig:arch}
\end{center}
\end{figure*}

Typically, singing involves more vocal effort and tends to have a slower rate of change of syllabic content along with enhanced higher frequency formants~\cite{sundberg1977acoustics,titze1992vocal}.  Due to the melodic nature of singing, a singer produces a smoother fundamental frequency
(F0) contour ~\cite{umbert2015expression}. 
The previous works on STS conversion can be broadly categorized into two directions : (i) Template-based STS  
\cite{cen2012template} 
and (ii) Model-based STS 
~\cite{saitou2007speech}. 
Both these methods require additional inputs like high quality singing templates or phoneme-score synchronization information, which are cumbersome in practice.

In this paper, inspired by the prior work by Parekh et. al. \cite{sp2si}, we consider the STS task as a style transfer problem. The proposed model uses a template of melody/pitch contour and poses the style transfer as a data driven learning problem. We also propose a novel neural network architecture for the style transfer where a supervised singing voice target with a suitable objective function is introduced.
Further, the adversarial loss function based on boundary equilibrium GAN~\cite{BEGAN}  is used for training the STS model. 

\begin{figure*}[t!]
\begin{center}
\centerline{\includegraphics[width=2.0\columnwidth]{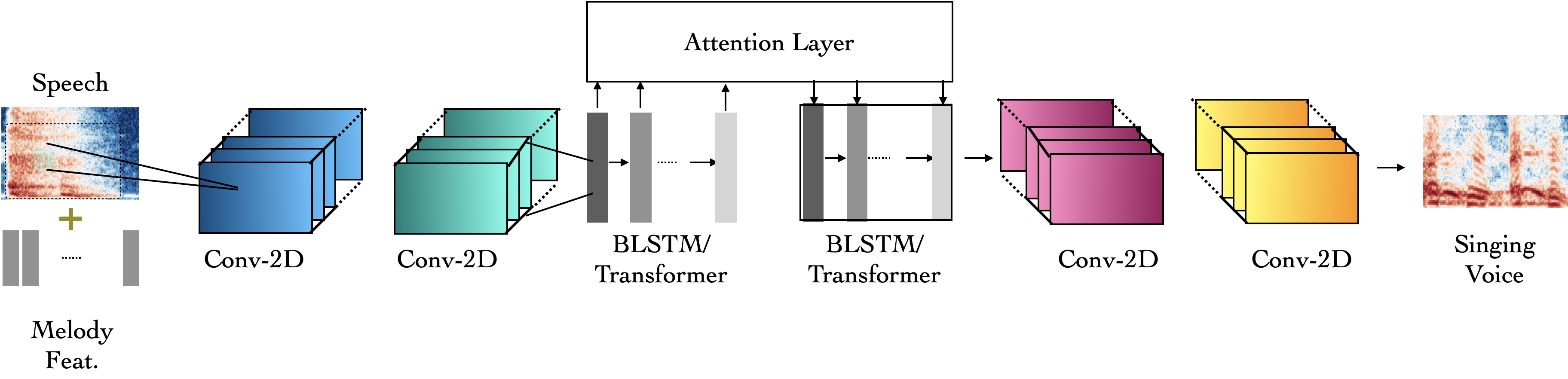}}
\vspace{-0.15in}
\caption{Proposed symmetrical convolutional recurrent network (SymNet) architecture with attention.}
\vspace{-0.3in}
\label{fig:symnet_arch}
\end{center}
\end{figure*}
The model architecture is similar to the encoder-decoder model~\cite{cho2014properties} popularly used in machine translation.
The proposed architecture is realized with a symmetric network (called SymNet) consisting of convolutional, transformer, and attention layers. The alignment between the speech and singing voice is achieved with the attentional layer between the central transformer layers while local signal characteristics are captured with convolutional layers at the beginning of encoder network as well as the end of decoder network. 
We also explore a data augmentation procedure using unpaired singing/speech data to mitigate the necessity of large amounts of paired data.
The experiments are performed on the NUS \cite{NUS} and the NHSS datasets \cite{sharmanhss}. In these experiments, we compare the proposed approach with prior state-of-the-work for STS. 
Using the proposed SymNet architecture, data augmentation and loss function, we observe significant performance improvements over baseline models in terms of the objective and subjective reconstruction quality.

\section{STS Framework}
The block schematic of the proposed model for STS is shown in Fig~\ref{fig:arch}. We represent the input speech with the log-magnitude spectrogram. A vocal melody extractor \cite{vocal_ext}, is used to extract melody contour from the reference/target singing voice. We then time-stretch the spectrogram of speech to the same length as the target F0 contour. The speech spectrogram are concatenated with the melody contour features and are  input to the deep neural model. 

\subsection{SymNet architecture}\label{sec:SymNet} 
The proposed SymNet architecture is shown in Fig~\ref{fig:symnet_arch}. 
The input speech spectrogram along with the melody features are passed to two  convolutional layers to model the local spectro-temporal patterns. The output of the convolutional layer is passed through a recurrent layer for modeling long-term dependencies. We consider bi-directional long-short term memory (BLSTM) or a transformer layers in this stage. The BLSTM model is a recurrent neural architecture implements using LSTM cells while the transformer models use self-attention layers to incorporate temporal dependency.  The next layer of processing involves a separate attention network which performs the non-linear time alignment between the speech and singing voice spectrogram based on the target melody features. The rest of the network consisting of a BLSTM/transformer layer and the two convolutional layers allows the architecture to form  a symmetrical model before and after the attention layer. 
We highlight the importance of the symmetrical architecture and the combination of three types of layers by comparing other possible architectures in our experiments.
The detailed SymNet configuration is summarized in Table \ref{SymNetConfig}. Each transformer layer has $4$ attention heads along with $2$ encoder and $2$ decoder layers. The dimension of the feed forward layer is set to $1024$.

\begin{table}[t!]
\caption{SymNet configuration. K: kernels, D - dimensions}
\label{SymNetConfig}
\vspace{-0.2in} 
\begin{center}
\begin{small}
\begin{sc}
\begin{tabular}{c | cc}
\hline
Layer & SymNet$_{256}$ & SymNet$_{512}$\\
\hline
Conv$\times2$ & 64K, $3\times3$ & 64K, $3\times3$\\  
BLSTM/Transformer & 256 D & 512 D \\  
Atten. & 256 D & 512 D \\
BLSTM/Transformer & 256 D & 512 D \\ 
Conv$\times2$ & 64K, $3\times3$ & 64K, $3\times3$ \\  
\hline
\end{tabular}
\end{sc}
\end{small}
\end{center}
\vspace{-0.4in}
\end{table}


\subsection{Data augmentation}\label{sec:DA} 
We utilize a data augmentation method that generates singing voice data from natural speech data. Our prior work \cite{our_work} used a data generation approach for automatic lyric recognition by leveraging the WORLD vocoder \cite{world12}. Specifically, the approach consisted of decomposing the  natural speech and singing voice (opera vocals) into three constituent components - spectral envelope, fundamental frequency (F0) and aperiodicity parameter. The F0 contour from the singing voice is then used along with the spectral envelope and the aperiodic parameter from the natural speech in the synthesizer. The synthesized output is the singing voice version of the natural speech. The western opera vocal dataset consists of both male and female opera singers and during the synthesis it was made sure that the speech and the opera vocals are gender and F0 matched~\cite{our_work}. 
We use the synthesized output and the input speech data as the paired data for augmentation. While this approach already generates a singing voice version of the natural speech, we find that the reconstruction quality is not sufficient for commercial applications. More details about the data augmentation can be found in  Basak et. al.~\cite{our_work}, 



\subsection{BEGAN loss}\label{sec:BEGAN}
The boundary equilibrium generative adversarial networks (BEGAN) ~\cite{berthelot2017began} is a energy based GAN, which matches the distribution of loss using an auto-encoder architecture, unlike the
original GAN which matches the distributions between the real and generated samples directly.
Moreover, the BEGAN relaxes the equilibrium of the auto-encoder loss using a hyper-parameter $\gamma \in [0,1]$. In order to implement the BEGAN loss ($L_{G}$)~\cite{BEGAN}, the STS model also has an additional discriminator network.  

\subsection{Training}\label{sec:train}
We experimented with two different training settings, (i) without the discriminator (and therefore without the BEGAN loss), and (ii) with the BEGAN discriminator. For the first one, our learning objective is based on mean square error ($L_{MSE}$) between predicted ($\hat{Y}$) and the true log-magnitude spectrograms of singing voice ($Y \in R^{F \times T}$). 
The combined loss then is, 
$L = L_{MSE} + \zeta L _G$. Here, $\zeta$ controls the trade-off between the MSE  and the GAN loss.

\subsection{Loss function annealing}
\label{sec:annealing}
We  explore a loss function annealing, where the regularization term $\zeta$ in the joint loss function is annealed as a function of the iteration to focus on the reconstruction MSE in the latter part of the training. 
In our experiments (Sec~\ref{sec:annealex}), we explored various annealing schedules, such as a linear decay and a step decay.


\section{Experiments}

\subsection{Dataset}
We use the NUS data \cite{NUS} as the paired training dataset containing $115$ mins of singing data and the corresponding $54$ mins of speech data. The $48$ recordings of $20$ unique English songs are sung/read by $12$ subjects, and each song-speech paired audio can be time-aligned with their phone annotations. The phone annotations are in accordance with the CMU dictionary ($39$ phonemes).
Out of the $20$ unique songs in the dataset, one song (with two recordings) is used as the test set for NUS (similar to the dataset division followed in \cite{sp2si}). We also use NHSS \cite{sharmanhss} for training/testing the models. It consist of $100$ songs with their respective speech recorded by $10$ singers, resulting in a total of $7$ hours of audio data. For data augmentation, the LibriSpeech \cite{panayotov2015librispeech} is converted to synthetic singing voice and about $10$ hours of augmented data is added for model training (Sec. \ref{sec:DA}). We use the log spectral distance (LSD) as the performance metric in our objective   evaluation. Further, a subjective test is performed to affirm the quality improvements seen in the objective metric.


\subsection{Input pre-processing}\label{sec:input}
Given an input time-domain speech signal and a target F0 contour, the pre-processing consists of the following steps:
\vspace{1mm}\\
\textbf{Silence frame removal} :\hspace{1mm} All the silence frames from the speech signal are removed. This is achieved via a short-time energy threshold set at $40$dB below the maximum energy frame. Any set of three or more consecutive silent frames (longer than approximately $50$ms) are removed.\vspace{1mm}\\
\textbf{Time stretching} :\hspace{1mm} The input spectrogram is linearly time-warped to the length of the F0 contour.\vspace{1mm}\\
    \textbf{Singing melody contour} :\hspace{1mm} The melody contour is extracted using the CREPE algorithm ~\cite{crepe} applied on the singing voice. We also convert the continuous-valued F0 to one of the $128$ MIDI levels. For training with paired data, we extract the melody contour from the singing counterpart of the input speech.\vspace{1mm}\\
    \textbf{Random re-sampling (RR)} :\hspace{1mm}  We have re-sampled the input speech in random fashion to change the rhythm and to disentangle the content and rhythm~\cite{BEGAN}. 

\begin{table}[t!]
\caption{Performance (LSD -dB) for the ablation study on the network architecture.   Here, L - Layers, C - Cells, A - Attention, and K - convolutional kernels. All models here are trained with MSE loss.}
\label{models}
\vspace{-0.2in}
\begin{center}
\begin{small}
\begin{sc}
\begin{tabular}{lccccr}
\hline
Arch. &Dim. & NUS & \textcolor{black}{NHSS}\\
\hline
BLSTM(3L)  & 256 & $10.45$ & \textcolor{black}{11.54} \\
BLSTM(3L) & 512 & $10.42$& \textcolor{black}{11.47} \\ \hline 
Conv(2L)-BLSTM(2L) &  256 & $10.38$ & \textcolor{black}{11.35} \\
Conv(2L)-BLSTM(2L) &  512 & $10.33$ & \textcolor{black}{11.31} \\ \hline 
BLSTM(2L)-A-BLSTM(2L) & 256 & $10.34$ & \textcolor{black}{11.02} \\
BLSTM(2L)-A-BLSTM(2L) & 512 & $10.29$ & \textcolor{black}{10.96} \\ \hline 
Conv(4L)-BLSTM(2L)-A &  256& $10.16$  & \textcolor{black}{10.88} \\
Conv(4L)-BLSTM(2L)-A &  512& $10.13$ & \textcolor{black}{10.85} \\
\hline 
\textcolor{black}{Conv(4L)-Trans.(2L)-A} & \textcolor{black}{ 256} & \textcolor{black}{9.99}  & \textcolor{black}{10.63} \\
\textcolor{black}{Conv(4L)-Trans.(2L)-A} & \textcolor{black}{ 512}& \textcolor{black}{9.89} & \textcolor{black}{10.56} \\
\hline
SymNet$_{256}$ (BLSTM) &  256& $10.02$ & \textcolor{black}{10.70} \\
SymNet$_{512}$ (BLSTM) &  512& ${9.95}$ & \textcolor{black}{10.61} \\
\hline
\textcolor{black}{SymNet$_{256}$(Trans.)} & \textcolor{black}{ 256}& \textcolor{black}{$9.93$} & \textcolor{black}{10.55} \\
\textcolor{black}{SymNet$_{512}$(Trans.)} & \textcolor{black}{ 512}& \textcolor{black}{$\textbf{9.87}$} & \textcolor{black}{\textbf{10.51}} \\
\hline
\end{tabular}
\end{sc}
\label{tab:tab1}
\end{small}
\end{center}
\vskip -0.1in
\end{table}

\begin{table}[t!]
\caption{Comparison of BEGAN loss annealing schedule for $\zeta$. The model architecture is SymNet$_{256}$ (BLSTM)  trained on NUS dataset.}
\label{tab:annealing}
\vspace{-0.2in}
\begin{center}
\begin{small}
\begin{sc}
\begin{tabular}{ccc}
\hline
Schedule for $\zeta$ & NUS  & \textcolor{black}{NHSS }\\
\hline
Constant $\zeta=0$ & $10.02$ &\textcolor{black}{10.70}\\
Constant $\zeta=0.3$ & $9.52$ &\textcolor{black}{10.41}\\
$\zeta=0.3$ decay by $0.001$ at each epoch & $9.45$ &\textcolor{black}{10.40}\\
First 15 epochs $\zeta=0.3$,  $\zeta=0$ after & $\textbf{9.33}$ &\textcolor{black}{\textbf{10.27}}\\

\hline
\end{tabular}
\end{sc}
\end{small}
\end{center}
\vspace {-0.15in} 
\end{table}
\begin{table*}[t!]
\caption{Comparison of the prior work with the proposed approach in terms of log spectral distance (LSD) in dB. Also reported are the incremental performance gains observed using various components. Here, Symnet corresponds to symmetric network with either the transformer or BLSTM layers.
}
\label{tab2}
\vspace{-0.2in}
\begin{center}
\begin{small}
\begin{sc}

 \begin{tabular}{lccccccr}
  \hline  
  
  \hline
Model & MSE loss & BEGAN loss & Data aug.  & w/ NHSS & LSD-NUS  & \textcolor{black}{LSD-NHSS}\\
 \hline
Parekh et. al.~\cite{sp2si} & \checkmark & & & & 11.22 & \textcolor{black}{11.91} \\  
Wu et al. \cite{BEGAN} & & \checkmark & &  & $9.6$ & \textcolor{black}{10.57}\\  
SymNet$_{512}$ (BLSTM) & \checkmark &  & &  & $9.95$  & \textcolor{black}{10.61}\\ 
SymNet$_{512}$ (Transformer) & \checkmark &  \checkmark & &  & $9.22$ & \textcolor{black}{10.13} \\ \hline
   
 SymNet$_{512}$ (BLSTM) &\checkmark &  \checkmark & \checkmark &  & $9.07$ & \textcolor{black}{9.98} \\ 
    \textcolor{black}{SymNet$_{512}$ (Transformer)} &\checkmark & \checkmark & \checkmark &  &  \textcolor{black}{$8.99$} & \textcolor{black}{9.62} \\ \hline
 SymNet$_{512}$ (BLSTM) &\checkmark &  \checkmark &   & \checkmark & \textcolor{black}{$9.01$}& \textcolor{black}{8.79} \\ 
  \textcolor{black}{SymNet$_{512}$ (Transformer)} &\checkmark & \checkmark & & \checkmark &  \textcolor{black}{8.81} & \textcolor{black}{8.74}\\ \hline
 SymNet$_{512}$ (BLSTM) &\checkmark &  \checkmark & \checkmark & \checkmark & \textcolor{black}{${8.80}$} & \textcolor{black}{8.76}\\
 \textcolor{black}{SymNet$_{512}$ (Transformer)}  &\checkmark &  \checkmark & \checkmark & \checkmark & \textcolor{black}{$\textbf{8.63}$} & \textcolor{black}{\textbf{8.67}}\\
 \hline
 

\hline
\end{tabular}
\end{sc}
\label{tab:StateOfArt} 
\end{small}
\end{center}
\vspace {-0.2in} 
\end{table*}

\subsection{Audio reconstruction}\label{sec:pred}
The magnitude specrogram is recovered by applying $g(\hat Y) = e^{\hat Y - 1}$  element-wise (the target spectrograms are ${\hat Y}=log(1+{\hat y})$, where $y$ is the magnitude spectrogram of the singing voice). The reconstructed magnitude spectrogram $g(\hat Y)$ is then converted to log-mel spectrogram with $80$ frequency bins. Then, the MelGAN vocoder \cite{kumar2019melgan} is used to generate audio waveforms from the spectrogram. 

\subsection{Implementation details}
The time domain signals are re-sampled from $44$kHz to $16$kHz. We compute short-time Fourier transform (STFT) with $1024$ frequency samples using $64$ ms window size and $16$ms hop size. 
For the first setting (the one with just the MSE loss), we use Adam optimizer~\cite{kingma2014adam} with initial learning rate of $0.003$ and an exponential decay factor of $0.9$. All the networks were trained for $30$ epochs, (with $1000$ mini-batches in each epoch). 
For the second setting, the one with BEGAN loss, we use Adam optimizer with initial learning rate of $0.001$ and exponential decay factor of $0.99$. All the neural network implementations   are performed using PyTorch~\cite{paszke2019pytorch} and Librosa~\cite{mcfee2015librosa}.

\subsection{Ablation study on architecture design}
\label{sec:results}
The first set of results, reported in Table~\ref{tab:tab1}, show the results for the ablation study with different model architecture choices. All the models reported here use the first setting of the loss function (only MSE). The  results reported here are for models with various architectures -  three BLSTM layers (with $256$ and $512$ cells), convolutional and BLSTM layers, BLSTM layers with attention, convolutional layers followed by BLSTM layers, convolutional and transformer layers and the proposed SymNet architectures. The key observations from there experiments  are,    
\begin{itemize}[leftmargin=*]
    \item The mixture of convolutional and transformer/LSTM layers is superior to the use of convolutional layers alone~\cite{BEGAN}. The use of $512$D is slightly better than $256$D.
    \item The use of attention in the model allows the non-linear time alignment between the speech and 
    singing voice counterparts with only a minor increase in the number of parameters.
    \item The transformer  improves over the BLSTM counterparts. 
    \item Combining all three types of layers (Conv-Trans-A) demonstrates the effectiveness of the three types of layers, namely, local modeling with convolution, long-term modeling with the transformer, and dynamic alignment with attention.
    \item SymNet improves the LSD over the models with the same set of layers and parameters, but are asymmetrical. To the best of our knowledge, this is the first attempt to highlight the importance of the symmetric architecture for audio style transfer tasks.  
\end{itemize}

\subsection{BEGAN loss annealing}\label{sec:annealex}
In Table~\ref{tab:annealing}, we consider three schedules for $\zeta$ which controls relative weighting of the BEGAN loss and the MSE loss - (i) $\zeta$ is fixed at $0.3$, (ii) $\zeta$ is initialized to $0.3$ and gradually reduced by $0.001$ for every epoch, (iii) $\zeta$ is initialized to $0.3$ and changed to 0 after $15$ epochs. Table \ref{tab:annealing} shows the results for the scheduling choices along with the system trained without BEGAN loss ($\zeta=0$). Even though BEGAN loss improves the LSD for fixed $\zeta$, annealing $\zeta$ further improves the performance. 

\subsection{Analysis of different components}
The results comparing different components of the proposed model are given in Table~\ref{tab:StateOfArt}.  The encoder-decoder model reported in \cite{BEGAN} serves as the baseline system. The results reported here show that the incorporation of the SymNet architecture with BEGAN loss improves over the baseline system.
Further, the incorporation of the data augmentation method improves the reconstruction quality. 
The effect of additional training data is explored by training the $SymNet_{512}$ model with additional data from NHSS dataset \cite{sharmanhss}. 
Further the combination of the data-augmentation approach along with the addition of the NHSS dataset improves the reconstruction quality with an absolute improvement of $0.97$ dB over the baseline system on the NUS dataset and $1.46$ dB on the NHSS dataset.

\begin{table}[t!]
\caption{Subjective evaluation (MOS) using hidden reference (ground truth) with $30$ subjects listening to $20$ audio snippets synthesized from the SymNet$_{512}$(Transformer) and the baseline system ( We et al.~\cite{BEGAN}). Both the systems used the same training data as well. }
\label{tab:sub_eval}
\vspace{-0.2in}
\begin{center}
\begin{small}
\begin{sc}
\begin{tabular}{cccc}
\hline
 & Hidden Ref. & Wu~\cite{BEGAN} & SymNet\\
\hline
MOS  & 4.86 $\pm 0.01$  & 2.71 $\pm 0.08$ & \textbf{3.08}  ${\boldsymbol \pm \textbf{0.09}}$\\
\hline
\end{tabular}
\end{sc}
\end{small}
\end{center}
\vspace {-0.2in} 
\end{table}

\subsection{Subjective Evaluation}
We conducted a mean opinion score (MOS) listening experiment\footnote{The study is hosted at \url{https://symnettesting.herokuapp.com/}} with $20$ audio clips, each with three versions - hidden reference (ground truth), style transferred from speech using the proposed SymNet approach and using the baseline system \cite{BEGAN}.
The three audio files for each snippet were presented in a random order. The participants belonged to the age group of $21-39$ with normal hearing.  The subjects were  asked to listen to the audio stimuli using high quality headphones and rate the files according to naturalness, reconstruction quality, intelligibility of the lyrics, and the preservation of speaker identity. The participants gave the rating on a $5$ point scale, with $1$ being extremely poor quality while $5$ represents the natural singing voice. In total, $30$ participants took part in the subjective listening test with $20$ audio snippets. The summary of the subjective listening tests results is given in Table.~\ref{tab:sub_eval}. The high rating for the hidden reference makes the results credible. The subjective results confirm the improved quality of the proposed SymNet model over the  baseline system.  The SymNet model improves the baseline system by an absolute MOS value of $0.37$. 
 
\section{Summary}
This paper proposes a novel architecture for the  task of audio style transfer where the objective is to convert the read speech to singing voice (STS). The proposed   architecture consists of a mixture of convolutional and transformer neural networks along with an attention layer. We   investigate additional novel components for data augmentation and loss function annealing. The proposed system achieves significant improvements in reconstruction quality over the stat-of-art approach using objective metrics.  Several experiments have been performed to analyze the incremental performance benefits for various components in the proposed system. Finally, subjective evaluations also confirm the superior performance of the proposed method. 


 
  





\ninept
\bibliographystyle{IEEEbib}
\bibliography{ref}

\begin{thebibliography}{10}

\bibitem{gatys2016image}
L.~A. Gatys, A.~S. Ecker, and M.~Bethge,
\newblock ``Image style transfer using convolutional neural networks,''
\newblock in {\em Proceedings of the IEEE conference on computer vision and
  pattern recognition}, 2016, pp. 2414--2423.

\bibitem{relate1}
A.~Haque, M.~Guo, and P.~Verma,
\newblock ``Conditional end-to-end audio transforms,''
\newblock {\em 10.21437/Interspeech.2018-38}, pp. 2295--2299, 09 2018.

\bibitem{relate2}
N.~Mor, L.~Wolf, A.~Polyak, and Y.~Taigman,
\newblock ``A universal music translation network,''
\newblock {\em arXiv preprint arXiv:1805.07848}, 2018.

\bibitem{relate3}
C.-W. Wu, J.-Y. Liu, Y.-H. Yang, and R.~J.-S. Jang,
\newblock ``Singing style transfer using cycle-consistent boundary equilibrium
  generative adversarial networks,''
\newblock {\em arXiv: arXiv:1807.02254}, 2018.

\bibitem{sundberg1977acoustics}
J.~Sundberg,
\newblock ``The acoustics of the singing voice,''
\newblock {\em Scientific American}, vol. 236, no. 3, pp. 82--91, 1977.

\bibitem{titze1992vocal}
I.~R. Titze and J.~Sundberg,
\newblock ``Vocal intensity in speakers and singers,''
\newblock {\em the Journal of the Acoustical Society of America}, vol. 91, no.
  5, pp. 2936--2946, 1992.

\bibitem{umbert2015expression}
M.~Umbert, J.~Bonada, M.~Goto, T.~Nakano, and J.~Sundberg,
\newblock ``Expression control in singing voice synthesis: Features,
  approaches, evaluation, and challenges,''
\newblock {\em IEEE Signal Processing Magazine}, vol. 32, no. 6, pp. 55--73,
  2015.

\bibitem{cen2012template}
L.~Cen, M.~Dong, and P.~Chan,
\newblock ``Template-based personalized singing voice synthesis,''
\newblock in {\em IEEE International Conference on Acoustics, Speech and Signal
  Processing (ICASSP)}, 2012, pp. 4509--4512.

\bibitem{saitou2007speech}
T.~Saitou, M.~Goto, M.~Unoki, and M.~Akagi,
\newblock ``Speech-to-singing synthesis: Converting speaking voices to singing
  voices by controlling acoustic features unique to singing voices,''
\newblock in {\em IEEE Workshop on Applications of Signal Processing to Audio
  and Acoustics}, 2007, pp. 215--218.

\bibitem{sp2si}
J.~{Parekh}, P.~{Rao}, and Y.~{Yang},
\newblock ``Speech-to-singing conversion in an encoder-decoder framework,''
\newblock in {\em IEEE International Conference on Acoustics, Speech and Signal
  Processing (ICASSP)}, 2020, pp. 261--265.

\bibitem{BEGAN}
D.-Y. Wu and Y.-H. Yang,
\newblock ``{Speech-to-Singing Conversion Based on Boundary Equilibrium GAN},''
\newblock in {\em Proc. Interspeech}, 2020, pp. 1316--1320.

\bibitem{cho2014properties}
K.~Cho, B.~Van~Merri{\"e}nboer, D.~Bahdanau, and Y.~Bengio,
\newblock ``On the properties of neural machine translation: Encoder-decoder
  approaches,''
\newblock {\em arXiv preprint arXiv:1409.1259}, 2014.

\bibitem{NUS}
Z.~{Duan}, H.~{Fang}, B.~{Li}, K.~C. {Sim}, and Y.~{Wang},
\newblock ``The {NUS} sung and spoken lyrics corpus: A quantitative comparison
  of singing and speech,''
\newblock in {\em Asia-Pacific Signal and Information Processing Association
  Annual Summit and Conference}, 2013, pp. 1--9.

\bibitem{sharmanhss}
B.~Sharma, X.~Gao, K.~Vijayan, X.~Tian, and H.~Li,
\newblock ``{NHSS}: A speech and singing parallel database,''
\newblock {\em arXiv preprint arXiv:2012.00337}, 2020.

\bibitem{vocal_ext}
L.~Su,
\newblock ``Vocal melody extraction using patch-based cnn,''
\newblock in {\em IEEE International Conference on Acoustics, Speech and Signal
  Processing (ICASSP)}, 2018, pp. 371--375.

\bibitem{our_work}
S.~Basak, S.~Agarwal, S.~Ganapathy, and N.~Takahashi,
\newblock ``End-to-end lyrics recognition with voice to singing style
  transfer,''
\newblock in {\em IEEE International Conference on Acoustics, Speech and Signal
  Processing (ICASSP)}, 2021.

\bibitem{world12}
M.~Morise, F.~Yokomori, and K.~Ozawa,
\newblock ``World: a vocoder-based high-quality speech synthesis system for
  real-time applications,''
\newblock {\em IEICE TRANSACTIONS on Information and Systems}, vol. 99, no. 7,
  pp. 1877--1884, 2016.

\bibitem{berthelot2017began}
D.~Berthelot, T.~Schumm, and L.~Metz,
\newblock ``Began: Boundary equilibrium generative adversarial networks,''
\newblock {\em arXiv preprint arXiv:1703.10717}, 2017.

\bibitem{panayotov2015librispeech}
V.~Panayotov, G.~Chen, D.~Povey, and S.~Khudanpur,
\newblock ``Librispeech: An {ASR} corpus based on public domain audio books,''
\newblock in {\em IEEE International Conference on Acoustics, Speech and Signal
  Processing (ICASSP)}, 2015, pp. 5206--5210.

\bibitem{crepe}
J.~Kim, J.~Salamon, P.~Li, and J.~Bello,
\newblock ``{CREPE}: A convolutional representation for pitch estimation,''
\newblock in {\em IEEE International Conference on Acoustics, Speech and Signal
  Processing (ICASSP)}, 2018.

\bibitem{kumar2019melgan}
K.~Kumar, R.~Kumar, T.~de~Boissiere, L.~Gestin, W.~Z. Teoh, J.~Sotelo,
  A.~de~Br{\'e}bisson, Y.~Bengio, and A.~Courville,
\newblock ``Melgan: Generative adversarial networks for conditional waveform
  synthesis,''
\newblock {\em arXiv preprint arXiv:1910.06711}, 2019.

\bibitem{kingma2014adam}
D.~P. Kingma and J.~Ba,
\newblock ``Adam: A method for stochastic optimization,''
\newblock {\em arXiv preprint arXiv:1412.6980}, 2014.

\bibitem{paszke2019pytorch}
A.~Paszke, S.~Gross, F.~Massa, A.~Lerer, J.~Bradbury, G.~Chanan, T.~Killeen,
  Z.~Lin, N.~Gimelshein, L.~Antiga, et~al.,
\newblock ``Pytorch: An imperative style, high-performance deep learning
  library,''
\newblock {\em arXiv preprint arXiv:1912.01703}, 2019.

\bibitem{mcfee2015librosa}
B.~McFee, C.~Raffel, D.~Liang, D.~P. Ellis, M.~McVicar, E.~Battenberg, and
  O.~Nieto,
\newblock ``Librosa: Audio and music signal analysis in {Python},''
\newblock in {\em Proceedings of the 14th python in science conference}, 2015,
  vol.~8, pp. 18--25.

\end{thebibliography}

\end{document}